\documentclass[aps,prl,twocolumn,groupedaddress,showpacs,reprint]{revtex4-2}
\usepackage{graphicx}
\usepackage{color}
\usepackage{amsmath}
\usepackage{bm}
\usepackage{txfonts}

\begin{document}

\title{
Irreversible efficiency and Carnot theorem for heat engines operating with multiple heat baths in linear response regime
}

\author{Yuki Izumida}
\affiliation{Department of Complexity Science and Engineering, Graduate School of Frontier Sciences, The University of Tokyo, Kashiwa 277-8561, Japan}
\thanks{izumida@k.u-tokyo.ac.jp}

\begin{abstract}
The Carnot theorem, one expression of the second law of thermodynamics, places a fundamental upper bound on the efficiency of heat engines operating between two heat baths.
The Carnot theorem can be stated in a more generalized form for heat engines operating with multiple heat baths, where the maximum efficiency is achieved for reversible heat engines operating quasistatically between two heat baths.
In this study, we determine the irreversible efficiency of heat engines operating with multiple heat baths in a linear response regime, i.e.,
under small temperature differences and a slow variation of the control parameters,
by quantifying the impact of the dissipation by irreversible operations.
The Carnot theorem is derived as a natural consequence of it.
Because the result obtained is based on the linear response relation and fluctuation-dissipation theorem in the universal framework of linear response theory, 
it has wide applicability to irreversible heat engines operating in the linear response regime.
\end{abstract}

\maketitle

About two centuries ago, Carnot in his celebrated study~\cite{C1824} 
revealed the existence of the fundamental upper bound of the efficiency of heat engines determined by the absolute temperatures of two heat baths, 
which led to the discovery of entropy by Clausius and provided the foundation for the formulation of thermodynamics.
For heat engines operating with two heat baths,
the Carnot theorem is formulated as~\cite{Ca1985}:
\begin{eqnarray}
\eta=\frac{W}{Q_H}\le 1-\frac{T_C}{T_H}\equiv \eta_{\rm C}\left(\frac{T_C}{T_H}\right),\label{eq.eta_carnot}
\end{eqnarray}
where $T_H$ and $T_C$ ($T_C<T_H$) are the absolute temperatures of the hot and cold heat baths, respectively, $Q_H$ is the heat input from the hot heat bath, $W$ is the work output, and $\eta$ is the efficiency of the heat energy conversion defined as the ratio of the work output to heat input.
The maximum value $\eta_{\rm C}$ is the Carnot efficiency, which is expressed as a function of the ratio of the absolute temperatures of the heat baths and is achieved for reversible heat engines such as a Carnot cycle.

A heat engine operating between two heat baths is a special cycle.
In general, we may consider general heat engine cycles operating with multiple heat baths; we may even consider heat engines with an infinite number of heat baths labeled with continuously changeable temperatures along the cycle.
The widely-known Carnot theorem Eq.~(\ref{eq.eta_carnot}) is then generalized as~\cite{K1968,note}
\begin{eqnarray}
\eta=\frac{W}{Q_{\rm in}} \le 1-\frac{T_{\rm min}}{T_{\rm max}}=\eta_{\rm C}\left(\frac{T_{\rm min}}{T_{\rm max}}\right),\label{eq.eta_max}
\end{eqnarray}
where $Q_{\rm in}$ represents the total heat input from the heat baths and $T_{\rm min}$ and $T_{\rm max}$ represent the minimum and maximum values of the temperatures of the heat baths along the cycle, respectively.
The equality of Eq.~(\ref{eq.eta_max}) is achieved for the reversible heat engines operating between two heat baths, 
in which case $T_{\rm min}=T_C$ and $T_{\rm max}=T_H$ hold~\cite{note}.
It should be noted that the reversibility condition alone does not yield the equality and the condition of the two heat baths is necessary.
This highlights the importance of the role of reversible heat engines operating under the isothermal condition.

Power, which is defined as the work per unit time, also characterizes practical heat engines~\cite{Ca1985}.
It is defined by
\begin{eqnarray}
P\equiv \frac{W}{\mathcal T},\label{eq.pow}
\end{eqnarray}
where $\mathcal T$ is the time required to complete one cycle.
Because the reversible heat engines operating quasistatically ($\mathcal T \to \infty$) produce vanishing power, 
they are of no practical use.
It is to be determined to what extent the efficiency of the irreversible heat engines operating in finite cycle time is reduced.
The Carnot theorem formulated in terms of the inequalities such as those presented in Eqs.~(\ref{eq.eta_carnot}) and (\ref{eq.eta_max}) do not provide any answers.
By quantifying the effect of the irreversible dissipation intrinsic in the finite-time operation of the heat engines based on non-equilibrium theories or specific models, various efficiency formulae for heat engines operating at maximum power~\cite{Y1957,C1957,N1958,VLF2014,MP2015,CA1975,VB2005,JH2007,SS2008,IO2009,EKLVB2010,BSC2011,IO2012,VB2013,IO2014,J2014,CPV2016,IO2017,CMG2017,APL2020,GAGMRCH2020}, trade-off relations between efficiency and power~\cite{BSS2015,RSP2016,BS2016,SST2016,PS2018,DS2018,HR2018,MIO2022,YMS2022}, and tighter bounds for efficiency than that provided by Eq.~(\ref{eq.eta_carnot})~\cite{BS2020,MM2020,MMPLG2021,HH2021,I2021,WM2022,FD2021,FD2021_2} have been obtained. 
So far, however, most of the studies have focused on the heat engines with two heat baths (i.e., the case represented in Eq.~(\ref{eq.eta_carnot})) or used different definitions of efficiency~\cite{BSS2015,BS2016,BS2020,MM2020,WM2022,I2021,FD2021,FD2021_2,MMPLG2021}, and less is known about the irreversible efficiency of the general heat engines to which the Carnot theorem represented in Eq.~(\ref{eq.eta_max}) can be applied.

In this study, we formulate the irreversible efficiency of general heat engines operating with multiple heat baths in a linear response regime
with explicit inclusion of the effect of dissipation. 
The Carnot theorem presented in Eq.~(\ref{eq.eta_max}) as inequality is naturally derived from it.
The linear response regime refers to a regime with small temperature differences (i.e., small $\Delta T\equiv T_{\rm max}-T_{\rm min}$) and a slow variation of the control parameters operating the heat engines.
The linear response theory is a universal framework for describing systems in non-equilibrium states in the vicinity of the equilibrium states~\cite{KTH1991,MPRV2008}.
It is established on the linear response relations, which connect the external forces applied to a system and their conjugate displacements
in terms of response functions.
The response function is related to the equilibrium time-correlation function by the fluctuation-dissipation theorem~\cite{KTH1991,MPRV2008}.
The result in our study is derived based only on the linear response theory applied to the heat engine cycles.
Because we do not assume any specific dynamics, working substance, and heat engine cycle, 
the formula derived has wide applicability to heat engines operating in the linear response regime.

Let $\Gamma$ be the phase-space variables that specify the state of a working substance (system) of a heat engine.
The heat engine is operated by a time-dependent parameter $\lambda(t)$ with a period $\mathcal T$, which is included in the Hamiltonian $H(\Gamma; \lambda(t))$ of the system.
The heat engine is in contact with the heat baths labeled with the absolute temperature $T(t)$, which changes periodically in time between $T_{\rm min}$ and $T_{\rm max}$ with the period $\mathcal T$.
Let $\mathcal P(\Gamma; \bm \Lambda(t))$ be a probability distribution function of $\Gamma$ at time $t$, where
$\bm \Lambda(t) \equiv (\lambda(t), T(t))$ denote the control parameters.

The heat engine undergoes a periodic change along the cycle.
For a sufficiently slow driving such that the quasistatic limit becomes a good approximation, 
at each instant of time along the cycle, the probability distribution is approximated by the canonical distribution described below:
\begin{eqnarray}
\mathcal P(\Gamma; \bm \Lambda(t))\simeq \mathcal P_{\rm eq}(\Gamma; \bm \Lambda(t))\equiv e^{-\frac{H(\Gamma; \lambda(t))-F(\bm \Lambda(t))}{k_{\mathrm B}T(t)}},\label{eq.equilibrium_distri}
\end{eqnarray}
where $F(\bm \Lambda(t))\equiv -k_{\mathrm B}T(t)\ln \int d\Gamma e^{-H(\Gamma; \lambda(t))/k_{\mathrm B}T(t)}$ and $k_{\mathrm B}$ are the Helmholtz free energy and Boltzmann constant, respectively.
In this description, the time $t$ merely denotes the label that differentiates the different equilibrium states.

For the energetics of the heat engine, we define the average energy of the system as:
\begin{eqnarray}
E(t)\equiv \left<H(\Gamma; \lambda(t))\right>_{\bm \Lambda(t)}=\int d\Gamma H(\Gamma; \lambda(t))\mathcal P(\Gamma; \bm \Lambda(t)),\label{eq.energy_av}
\end{eqnarray}
where $\left<\cdot \right>_{\bm \Lambda(t)}$ denotes an average with respect to $\mathcal P(\Gamma; \bm \Lambda(t))$.
By differentiating $E(t)$ with respect to $t$ as $\dot E=\frac{d}{dt}\int d\Gamma H(\Gamma; \lambda(t))\mathcal P(\Gamma; \bm \Lambda(t))$,
where the overdot denotes the time derivative, we can decompose $E(t)$ into the sum of work and heat fluxes $j_w(t)$ and $j_q(t)$, described as follows: 
\begin{eqnarray}
\dot E&&=\int d\Gamma \frac{\partial H}{\partial t}\mathcal P(\Gamma; \bm \Lambda(t))+\int d\Gamma H(\Gamma; \lambda(t))\frac{\partial \mathcal P(\Gamma; \lambda(t))}{\partial t}\nonumber\\
&&=\left<\frac{\partial H(\Gamma; \lambda(t))}{\partial \lambda}\right>_{\bm \Lambda(t)} \dot{\lambda}+\int d\Gamma H(\Gamma; \lambda(t)) \frac{\partial \mathcal P(\Gamma; \bm \Lambda(t))}{\partial \bm \Lambda}\cdot \dot{\bm \Lambda}\nonumber\\
&&\equiv -j_w(t)+j_q(t).\label{eq.energy}
\end{eqnarray}
The work flux $j_w(t)$ can be written in terms of thermodynamic variables as
\begin{eqnarray}
j_w(t)=\left<p\right>_{\bm \Lambda(t)}\dot \lambda,\label{eq.jw}
\end{eqnarray}
where we have defined the generalized pressure $p\equiv -\partial H/\partial \lambda$.
We can also represent $j_q(t)$ in terms of thermodynamic variables.
We introduce the Shannon entropy of the system:
\begin{eqnarray}
\left<s\right>_{\bm \Lambda(t)}=-k_{\mathrm B} \int d\Gamma \mathcal P(\Gamma; \bm \Lambda(t))\ln \mathcal P(\Gamma; \bm \Lambda(t)),\label{eq.entropy}
\end{eqnarray}
where $s\equiv -k_{\mathrm B}\ln \mathcal P(\Gamma; \bm \Lambda(t))$ is the stochastic entropy~\cite{U2012,PP2021}.
For a sufficiently slow driving where Eq.~(\ref{eq.equilibrium_distri}) holds, we can represent the heat flux $j_q(t)$ in terms of the equilibrium entropy
$S\equiv \left<s\right>_{{\rm eq}, \bm \Lambda(t)}$ and the temperature $T(t)$ as
\begin{eqnarray}
j_q(t)\simeq T(t) \dot S,\label{eq.jq}
\end{eqnarray}
where $\left<\cdot \right>_{{\rm eq}, \bm \Lambda(t)}$ is the average with respect to $\mathcal P_{\rm eq}(\Gamma; \bm \Lambda(t))$.
It should be noted that the expression of $j_q(t)$ in terms of $S$ is valid only for the near-equilibrium condition in contrast to Eq.~(\ref{eq.jw}).
The expressions Eqs.~(\ref{eq.jw}) and (\ref{eq.jq}) as the product of the conjugate thermodynamic variables are convenient for the application of the linear response theory to the heat engine cycle below.

For further evaluation of $j_w(t)$ and $j_q(t)$ toward the formulation of the efficiency of the heat engine cycle, we recall the linear response theory.
Let $\bm \Lambda=(\lambda, T)$ be the control parameters of any point on the cycle as a reference state whose small variation can be considered as a small perturbation to the system. The entire cycle is constructed as an accumulation of such small variations.

First, we consider the static response of thermodynamic variables when a perturbation is added quasistatically.
Upon the introduction of a small variation in the control parameters $\bm \Lambda \to \bm \Lambda+\delta \bm \Lambda$, 
where $\delta \bm \Lambda \equiv (\delta \lambda,\delta T)$ serve as the external forces, the canonical distribution Eq.~(\ref{eq.equilibrium_distri}) changes as~\cite{KTH1991}:
\begin{eqnarray} 
&&\mathcal P_{\rm eq}(\Gamma; \bm \Lambda+\delta \bm \Lambda)\simeq \mathcal P_{\rm eq}(\Gamma; \bm \Lambda)\left(1+\frac{1}{k_{\mathrm B}T}\Delta \bm A \cdot \delta \bm \Lambda \right).\ \ \ \ \label{eq.equilibrium_distri_2}
\end{eqnarray}
Here, $\bm A$ is the conjugate displacement to $\delta \bm \Lambda$ represented as
\begin{eqnarray}
\bm A=(A_\lambda, A_T)\equiv (p, s)=\left(-\frac{\partial H}{\partial \lambda}, -k_{\mathrm B}\ln \mathcal P\right),\label{eq.Delta_A}
\end{eqnarray}
where $\Delta \bm A \equiv \bm A-\left<\bm A \right>_{{\rm eq}, \bm \Lambda}$, and we used $\Delta s \simeq \Delta H/T$ for the derivation of Eq.~(\ref{eq.equilibrium_distri_2}), which is shown from Eqs.~(\ref{eq.equilibrium_distri}), (\ref{eq.energy_av}), and (\ref{eq.entropy}).
It should be noted that $\left<\Delta \bm A\right>_{{\rm eq}, \bm \Lambda}=\bm 0$ from the definition.
By using Eqs.~(\ref{eq.equilibrium_distri_2}) and (\ref{eq.Delta_A}), we can derive $(\mu, \nu=\lambda, T)$
\begin{eqnarray}
\left<A_\mu \right>_{{\rm eq}, \bm \Lambda+\delta \bm \Lambda}=\left<A_\mu \right>_{{\rm eq}, \bm \Lambda}+\chi_{\mu \nu} \delta \Lambda_\nu,\label{eq.DA_average}
\end{eqnarray}
where $\left<\cdot \right>_{{\rm eq}, \bm \Lambda+\delta \bm \Lambda}$ is the average with respect to $\mathcal P_{\rm eq}(\Gamma; \bm \Lambda+\delta \bm \Lambda)$ and
\begin{eqnarray}
\chi_{\mu \nu} \equiv \frac{\partial \left<A_\mu \right>_{{\rm eq}, \bm \Lambda}}{\partial \Lambda_\nu}\label{eq.static_res}
\end{eqnarray}
denotes the static response coefficient.
The symmetry of the static response coefficients can be shown from the Maxwell relation:
\begin{eqnarray}
\chi_{\mu \nu}=\frac{\partial \left<A_{\mu}\right>_{{\rm eq}, \bm \Lambda}}{\partial \Lambda_\nu}=-\frac{\partial^2 F}{\partial \Lambda_\mu \partial \Lambda_\nu}=\frac{\partial \left<A_{\nu}\right>_{{\rm eq}, \bm \Lambda}}{\partial \Lambda_\mu}=\chi_{\nu \mu},
\end{eqnarray}
where we used $\left<A_\mu \right>_{{\rm eq}, \bm \Lambda}=-\partial F/\partial \Lambda_\mu$.

In general, a perturbation is added dynamically in a time-dependent manner.
We generalize the above static response of $\bm A$ so that it is included as a special case.
Let us consider that the system in the equilibrium state with $\bm \Lambda(t-\delta t)=(\lambda(t-\delta t), T(t-\delta t))$ at time $t-\delta t$ 
is perturbed by the time-dependent variation $\delta \bm \Lambda(t')\equiv \bm \Lambda(t')-\bm \Lambda(t-\delta t)$ until $t'=t$, where $\delta \bm \Lambda(t-\delta t)=\bm 0$.
Here, the time increment $\delta t$ is chosen such that it is sufficiently small in a macroscopic time scale while it is sufficiently large compared to a microscopic time scale, such as the correlation time of the system.
Following the standard linear response theory~\cite{KTH1991}, we can describe the linear response of the thermodynamic variables $\bm A$ of the system upon the addition of this perturbation as:
\begin{eqnarray}
\left<\Delta A_\mu \right>_{\bm \Lambda(t)}&&=\left<A_\mu \right>_{\bm \Lambda(t)}-\left<A_\mu \right>_{{\rm eq}, \bm \Lambda(t-\delta t)}\nonumber\\
&&=\int_{t-\delta t}^t \Phi_{\mu \nu}(t-t')\delta \Lambda_\nu (t')dt',\ \ \ \ \ \ \label{eq.response}
\end{eqnarray}
where $\Phi_{\mu \nu}$ denotes the response function satisfying the causality $\Phi_{\mu \nu}(t-t')=0$ for $t<t'$. 
The response function is related to the equilibrium time-correlation function by the fluctuation-dissipation theorem:
\begin{eqnarray}
\Phi_{\mu \nu}(\tau)=-\frac{\theta(\tau)}{k_{\mathrm B}T}\frac{d}{d\tau}C_{\mu \nu}(\tau),\label{eq.response_func}
\end{eqnarray}
where $\theta(\tau)$ is the Heviside step function and $C_{\mu \nu}(\tau)$ is the equilibrium time-correlation function between $\Delta A_\mu$ and $\Delta A_\nu$ defined by
\begin{eqnarray}
C_{{\mu \nu}, \bm \Lambda(t-\delta t)}(\tau)
\equiv \left<\Delta A_\mu(\tau)\Delta A_\nu(0)\right>_0.
\end{eqnarray}
Here, $\left<\right>_0$ denotes the average with respect to the unperturbed equilibrium dynamics with $\bm \Lambda$ being held fixed as 
$\bm \Lambda(t-\delta t)$. In particular, $C_{{\mu \nu}, \bm \Lambda(t-\delta t)}(0)=\left<\Delta A_\mu(0) \Delta A_\nu(0) \right>_0=\left<\Delta A_\mu \Delta A_\nu \right>_{{\rm eq}, \bm \Lambda(t-\delta t)}$ is the equilibrium correlation function.
By changing the variable from $t'$ to $\tau=t-t'$ in Eq.~(\ref{eq.response}), we have $\int_{t-\delta t}^t \Phi_{\mu \nu}(t-t')\delta \Lambda_\nu (t')dt'=\int_0^{\delta t} \Phi_{\mu \nu}(\tau)\delta \Lambda_\nu (t-\tau)d\tau$.
The range of $\tau$ that contributes to the integral is less than the correlation time that characterizes $C_{{\mu \nu}, \bm \Lambda(t-\delta t)}(\tau)$~\cite{KTH1991}, which is sufficiently smaller than $\delta t$.
We can thus replace the upper limit of the integral with infinity as $\int_0^{\delta t} \Phi_{\mu \nu}(\tau)\delta \Lambda_\nu (t-\tau)d\tau \simeq \int_0^{\infty} \Phi_{\mu \nu}(\tau)\delta \Lambda_\nu (t-\tau)d\tau$.
Moreover, because we are considering a sufficiently slow change of the control parameters, 
the change in $\delta \Lambda_\nu$ during the correlation time is sufficiently small, and 
we can expand $\Lambda_\nu(t-\tau)$ in terms of $\tau$ as $\delta \Lambda_\nu(t-\tau)=\delta \Lambda_\nu(t)-\dot \Lambda_\nu(t) \tau+O(\tau^2)$,
where we used $\delta \dot \Lambda_\nu(t')=\dot \Lambda_\nu(t')$.
By using this approximation of $\delta \Lambda_\nu(t-\tau)$ and Eq.~(\ref{eq.response_func}), we obtain:
\begin{eqnarray}
&&\int_0^\infty \Phi_{\mu \nu}(\tau)\delta \Lambda_\nu (t-\tau)d\tau
\simeq \chi_{{\mu \nu}, \bm \Lambda(t-\delta t)} \delta \Lambda_\nu (t)-R_{{\mu \nu}, \bm \Lambda(t-\delta t)}{\dot \Lambda}_v(t),\nonumber\\
\label{eq.response_approx}
\end{eqnarray}
where we used partial integration with respect to $\tau$ and $C_{{\mu \nu}, \bm \Lambda(t-\delta t)}(\infty)=0$, and assumed $\lim_{\tau \to \infty} C_{{\mu \nu}, \bm \Lambda(t-\delta t)}(\tau)\tau=0$.
Here, the static response coefficient $\chi_{\mu \nu}$ in Eq.~(\ref{eq.static_res}) is expressed in terms of the correlation function (Kirkwood relation):
\begin{eqnarray}
\chi_{{\mu \nu}, \bm \Lambda(t-\delta t)}=\frac{1}{k_{\mathrm B}T}C_{{\mu \nu}, \bm \Lambda(t-\delta t)}(0),
\end{eqnarray}
where it is a positive semi-definite symmetric matrix, as evidenced from their construction using the covariance matrix.
We also defined the generalized friction coefficient~\cite{PP2021}:
\begin{eqnarray}
R_{{\mu \nu}, \bm \Lambda(t-\delta t)}\equiv \frac{1}{k_{\mathrm B}T}\int_0^\infty C_{{\mu \nu}, \bm \Lambda(t-\delta t)}(\tau)d\tau,
\end{eqnarray}
which is a positive semi-definite matrix, reflecting the stability of equilibrium dynamics.
The symmetry is also derived from the time-reversal and time-translation invariance of the equilibrium dynamics for time-reversible symmetric observables at equilibrium, such as the thermodynamic variables under consideration~\cite{PP2021}.
From Eqs.~(\ref{eq.response}) and (\ref{eq.response_approx}), we have
\begin{eqnarray}
\left<A_\mu \right>_{\bm \Lambda(t)} && \simeq \left<A_\mu \right>_{{\rm eq}, \bm \Lambda(t-\delta t)}+\chi_{{\mu \nu}, \bm \Lambda(t-\delta t)} \delta \Lambda_\nu(t)-R_{{\mu \nu}, \bm \Lambda(t-\delta t)} \dot{\Lambda}_\nu(t)\nonumber\\
&&=\left<A_\mu \right>_{{\rm eq}, \bm \Lambda(t)}-R_{{\mu \nu}, \bm \Lambda(t-\delta t)}\dot{\Lambda}_\nu(t)\nonumber\\
&&\simeq \left<A_\mu \right>_{{\rm eq}, \bm \Lambda(t)}-R_{{\mu \nu}, \bm \Lambda(t)}\dot{\Lambda}_\nu(t),\label{eq.A_response}
\end{eqnarray}
where we used Eq.~(\ref{eq.DA_average}) in the second equality and approximated as $R_{{\mu \nu}, \bm \Lambda(t-\delta t)}\simeq R_{{\mu \nu}, \bm \Lambda(t)}$ in the third equality.

\begin{figure}
\begin{center}
\includegraphics[scale=0.3]{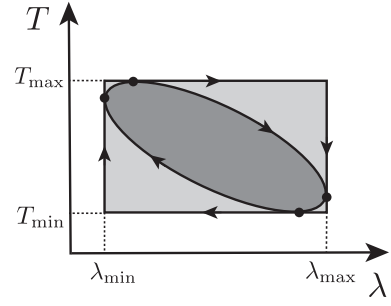}
\caption{Schematic illustration of the heat engine cycle in the space of the control parameters $\bm \Lambda=(\lambda, T)$.
The outer rectangular-shaped cycle represents an idealized cycle, which attains the upper bound in Eq.~(\ref{eq.main_eta}) (see the main text).}\label{enginecycle}
\end{center}
\end{figure}
By using Eq.~(\ref{eq.A_response}), we can construct the heat engine cycle (Fig.~\ref{enginecycle}) and formulate its efficiency.
We assume the following specific form as $T(t)$:
\begin{eqnarray}
T(t)\equiv T_{\rm min}+\gamma_q(t)\Delta T,\label{eq.T}
\end{eqnarray}
where $\gamma_q(t)$ is an arbitrary periodic function satisfying $0\le \gamma_q(t)\le 1$ and $\Delta T=T_{\rm max}-T_{\rm min}$.
Meanwhile, we assume that the protocol $\lambda(t)$ satisfies the following condition:
\begin{eqnarray}
&&\lambda(0)=\lambda(\mathcal T)=\lambda_{\rm min},\ \lambda(\mathcal T_{\rm in})=\lambda_{\rm max},
\end{eqnarray}
where the working substance absorbs heat during $0\le t \le \mathcal T_{\rm in}$ and rejects heat during $\mathcal T_{\rm in}< t \le \mathcal T$.
It should be noted that $T(0)\ne T_{\rm min}$ ($\gamma_q(0) \ne 0$) and $T(\mathcal T_{\rm in})\ne T_{\rm max}$ ($\gamma_q(\mathcal T_{\rm in}) \ne 1$) in general.

The efficiency of the cycle is obtained as:
\begin{eqnarray}
\eta=\frac{W}{Q_{\rm in}}=\frac{\int_0^{\mathcal T}j_w(t)dt}{\int_0^{\mathcal T_{\rm in}}j_q(t)dt}\label{eq.def_eta}.
\end{eqnarray}
We wish to evaluate this quantity up to the lowest order of the small quantities.
By using Eqs.~(\ref{eq.jw}), (\ref{eq.A_response}), and (\ref{eq.T}), 
we can approximate $W$ in Eq.~(\ref{eq.def_eta}) as: 
\begin{eqnarray}
W\simeq \int_0^{\mathcal T} \chi_{{\lambda T}, (\lambda(t), T_{\rm min})}\gamma_q(t)\dot \lambda dt \Delta T-\int_0^{\mathcal T} R_{{\lambda \lambda}, (\lambda(t), T_{\rm min})}\dot \lambda^2dt.\ \ \ \ \
\end{eqnarray}
Here, we used the approximation $\left<p \right>_{{\rm eq}, \bm \Lambda(t)}\simeq \left<p \right>_{{\rm eq}, (\lambda(t), T_{\rm min})}+\chi_{{\lambda T}, (\lambda(t), T_{\rm min})}\gamma_q(t) \Delta T$, where $\oint \left<p\right>_{{\rm eq}, (\lambda, T_{\rm min})}d\lambda=-\oint \partial F(\lambda, T_{\rm min})/\partial \lambda d\lambda=0$, and neglected the higher-order term of $O(\dot \lambda \dot \gamma_q \Delta T, \dot \lambda^2 \Delta T)$.
Because $W$ in the numerator is constituted with the small quantities, it is sufficient to evaluate $Q_{\rm in}$ in the denominator up to $O(1)$: 
\begin{eqnarray}
Q_{\rm in}\simeq T_{\rm min}(S(\lambda_{\rm max}, T_{\rm min})-S(\lambda_{\rm min}, T_{\rm min})),\ \ \ \ \ \ \label{eq.Qin_approx}
\end{eqnarray}
where we used Eqs.~(\ref{eq.jq}) and (\ref{eq.T}).
Using Eqs.~(\ref{eq.def_eta})--(\ref{eq.Qin_approx}) and the Maxwell relation $\chi_{{\lambda T}, (\lambda(t), T_{\rm min})}=\chi_{{T \lambda}, (\lambda(t), T_{\rm min})}=\partial S/\partial \lambda|_{\bm \Lambda=(\lambda(t), T_{\rm min})}$ and reparameterizing the time-dependent quantities as $\lambda(t)=\tilde \lambda(\theta)$, $\gamma_q(t)=\tilde \gamma_q(\theta)$, $S(\bm \Lambda(t))=\tilde S(\tilde{\bm \Lambda}(\theta))$, and $R_{\lambda \lambda, \bm \Lambda(t)}=\tilde R_{\lambda \lambda, \tilde{\bm \Lambda}(\theta)}$ in terms of $\theta \equiv t/\mathcal T$ ($0\le \theta \le 1$), 
we finally obtain:
\begin{eqnarray}
\eta&&=\eta_{\rm rev}-\frac{\int_0^1 \tilde R_{{\lambda \lambda}, (\tilde \lambda(\theta), T_{\rm min})}\left(\frac{d\tilde \lambda}{d\theta}\right)^2d\theta}{T_{\rm min}(S(\lambda_{\rm max}, T_{\rm min})-S(\lambda_{\rm min}, T_{\rm min}))\mathcal T}.\label{eq.main_eta}
\end{eqnarray}
Here, the first term
\begin{eqnarray}
\eta_{\rm rev}\equiv \frac{\int_0^1 \frac{\partial \tilde S(\tilde \lambda(\theta), T_{\rm min})}{\partial \theta}\tilde \gamma_q(\theta)d\theta}{S(\lambda_{\rm max}, T_{\rm min})-S(\lambda_{\rm min}, T_{\rm min})}\eta_{\rm C}\left(\frac{T_{\rm min}}{T_{\rm max}}\right)\label{eq.eta_rev}
\end{eqnarray}
denotes the reversible efficiency with $\eta_{\rm C}(T_{\rm min}/T_{\rm max})\simeq \Delta T/T_{\rm min}$ and the non-positive second term denotes the dissipation effect due to the finite-time operation of heat engines.
$\eta_{\rm rev}$ is achieved for the quasistatic limit $\mathcal T \to \infty$.
As $0 \le \tilde \gamma_q \le 1$ and $\partial \tilde S/\partial \theta \ge 0$ for $0\le \theta \le \theta_{\rm in}\equiv \mathcal T_{\rm in}/\mathcal T$ and $\partial \tilde S/\partial \theta \le 0$ for $\theta_{\rm in}< \theta \le 1$, we have $\eta_{\rm rev}\le \eta_{\rm C}(T_{\rm min}/T_{\rm max})$.
The equality is achieved for the heat engine cycles, 
in which the heat input upon the variation of $\lambda$ is operated under the isothermal condition.
In the space of the control parameters $\bm \Lambda=(\lambda, T)$, such an idealized cycle is given, e.g., as a rectangular-shaped cycle constituted with the isothermal processes with $\lambda_{\rm min}\to \lambda_{\rm max}$ at $T=T_{\rm max}$ ($\gamma_q=1$) and $\lambda_{\rm max}\to \lambda_{\rm min}$ at $T=T_{\rm min}$ ($\gamma_q=0$), connected by the isochoric processes with $T_{\rm max}\to T_{\rm min}$ at $\lambda=\lambda_{\rm max}$ and $T_{\rm min}\to T_{\rm max}$ at $\lambda=\lambda_{\rm min}$ (Fig.~\ref{enginecycle})~\cite{note2}.
Equation~(\ref{eq.main_eta}) together with Eq.~(\ref{eq.eta_rev}) constitute our main result.

A few remarks are in order with respect to the main result Eqs~(\ref{eq.main_eta}) and (\ref{eq.eta_rev}).
First, the Carnot theorem for heat engines operating with multiple heat baths in Eq.~(\ref{eq.eta_max}) is derived from Eq.~(\ref{eq.main_eta}): 
As the second term of the right-hand side of Eq.~(\ref{eq.main_eta}) is non-positive, we obtain $\eta \le \eta_{\rm rev}$. Moreover, as $\eta_{\rm rev}\le \eta_{\rm C}(T_{\rm min}/T_{\rm max})$, as stated above, Eq.~(\ref{eq.eta_max}) can be derived.

Furthermore, by choosing the optimal protocol for the parameter $\lambda(t)$ for a fixed cycle time, we can obtain another bound for $\eta$ as:
\begin{eqnarray}
\eta \le \eta_{\rm rev}-\frac{\mathcal L^2}{T_{\rm min}(S(\lambda_{\rm max})-S(\lambda_{\rm min}))\mathcal T},\label{eq.eta_bound_opt}
\end{eqnarray}
where $\mathcal L$ is the minimum value of $\Big[\int_0^1 \tilde R_{{\lambda \lambda}, (\tilde \lambda(\theta), T_{\rm min})}\left(\frac{d\tilde \lambda}{d\theta}\right)^2d\theta \Big]^{1/2}$ and is called the thermodynamic length~\cite{PP2021,SB1983}:
\begin{eqnarray}
\mathcal L &&\equiv \left|\int_0^1 \sqrt{\tilde R_{{\lambda \lambda}, (\tilde \lambda(\theta), T_{\rm min})}\left(\frac{d\tilde \lambda}{d\theta}\right)^2} d\theta \right| \nonumber\\
&&\le \sqrt{\int_0^1 \tilde R_{{\lambda \lambda}, (\tilde \lambda(\theta), T_{\rm min})} \left(\frac{d\tilde \lambda}{d\theta}\right)^2d\theta}. \  \  \  \
\end{eqnarray}
Here, we used the Cauchy--Schwartz inequality, and the equality is achieved for the optimal protocol, which makes the integrand a constant without any dependence on $\theta$.
The thermodynamic length is used to describe the minimum dissipation during near-equilibrium transitions~\cite{SB1983,SS1997,S2010,C2007,SC2012,ZSCD2012,BD2014,ZD2015,DB2020,DB2020,AMPLM2020,AAPLA2022} and is invariant under re-parameterization.
Equation~(\ref{eq.main_eta}) is tighter than $\eta \le \eta_{\rm rev}$ as it includes the extent of dissipation and cycle time explicitly. 
Similar forms of Eq.~(\ref{eq.eta_bound_opt}) have been obtained recently for adiabatically driven heat engines using different definitions of efficiency~\cite{BS2020,MM2020,WM2022,I2021,FD2021,FD2021_2,MMPLG2021}.

We can also obtain the efficiency at maximum power by using Eq.~(\ref{eq.main_eta}).
The power in Eq.~(\ref{eq.pow}) is obtained by:
\begin{eqnarray}
&&P \simeq \frac{\int_0^1 \frac{\partial \tilde S(\tilde \lambda(\theta), T_{\rm min})}{\partial \theta}\tilde \gamma_q(\theta)d\theta \Delta T}{\mathcal T}-\frac{\int_0^1 \tilde R_{{\lambda \lambda}, (\tilde \lambda(\theta), T_{\rm min})}\left(\frac{d\tilde \lambda}{d\theta}\right)^2d\theta}{\mathcal T^2}.\ \ \ \ \ \ \label{eq.def_pow}
\end{eqnarray}
By maximizing $P$ with respect to $\mathcal T$, we obtain the maximum power as
\begin{eqnarray}
P^*=\frac{\Big[\int_0^1 \frac{\partial \tilde S(\tilde \lambda(\theta), T_{\rm min})}{\partial \theta}\tilde \gamma_q(\theta)d\theta \Big]^2}{4\int_0^1 \tilde R_{{\lambda \lambda}, (\tilde \lambda(\theta), T_{\rm min})}\left(\frac{d\tilde \lambda}{d\theta}\right)^2d\theta}\Delta T^2.\label{eq.power_max}
\end{eqnarray}
Here, the cycle time at maximum power is $\mathcal T^*=2\int_0^1 \tilde R_{{\lambda \lambda}, (\tilde \lambda(\theta), T_{\rm min})}\left(\frac{d\tilde \lambda}{d\theta}\right)^2d\theta/\int_0^1 \frac{\partial \tilde S(\tilde \lambda(\theta), T_{\rm min})}{\partial \theta}\tilde \gamma_q(\theta)d\theta \Delta T$. 
By using Eq.~(\ref{eq.main_eta}) and $\mathcal T^*$, we obtain the efficiency at maximum power $\eta^*$:
\begin{eqnarray}
\eta^*=\frac{\eta_{\rm rev}}{2}\le \frac{1}{2}\eta_{\rm C}\left(\frac{T_{\rm min}}{T_{\rm max}}\right).\label{eq.effi_pmax}
\end{eqnarray}
Eq.~(\ref{eq.effi_pmax}) states that the efficiency of any heat engine cycle operating at maximum power in the linear response regime is bounded from above by one half of the Carnot efficiency, where the equality condition agrees with that of $\eta_{\rm rev}\le \eta_{\rm C}(T_{\rm min}/T_{\rm max})$.
The inequality Eq.~(\ref{eq.effi_pmax}) shows the advantages of heat engines operating with two heat baths in the context of powerful heat engines.

As a final remark, we note that the reversible efficiency $\eta_{\rm rev}$ can be derived in terms of equilibrium thermodynamics.
A reversible heat engine cycle can be approximated as an infinite sum of infinitesimally-small reversible Carnot cycles.
For each infinitesimally-small Carnot cycle operating between $T_{\rm min}$ and $T_{\rm min}+\delta T$, 
the infinitesimal work and heat are approximated as $\delta W\simeq \delta p\delta \lambda=(\partial p/\partial T)_\lambda \delta T\delta \lambda$ and $\delta Q\simeq (T_{\rm min}+\delta T)(\partial S/\partial \lambda)_T \delta \lambda$, respectively. The efficiency of this infinitesimally-small cycle is the Carnot efficiency:
$\eta=\delta W/\delta Q=\delta T/(T_{\rm min}+\delta T) \simeq \delta T/T_{\rm min}$,
where we used the Maxwell relation $(\partial p/\partial T)_\lambda=(\partial S/\partial \lambda)_T$. By stretching this cycle into the direction of $\lambda$ and placing $\delta T=\delta \tilde T(\theta)=\tilde \gamma_q(\theta) \Delta T$ parameterized by $\theta$, we obtain
\begin{eqnarray}
\eta=\frac{\oint \delta W}{\int \delta Q}\simeq \frac{\int_0^1 \frac{\partial \tilde S(\tilde \lambda(\theta), T_{\rm min})}{\partial \theta}\tilde \gamma_q(\theta) d\theta}{S(\lambda_{\rm max}, T_{\rm min})-S(\lambda_{\rm min}, T_{\rm min})}\frac{\Delta T}{T_{\rm min}},
\end{eqnarray}
where we used $(\partial \tilde S(\tilde \lambda, T_{\rm min})/\partial \tilde \lambda)_T(d\tilde \lambda/d\theta)=\partial \tilde S(\tilde \lambda(\theta), T_{\rm min})/\partial \theta$.
Thus, we recover the reversible efficiency of a heat engine cycle characterized by $\gamma_q$, as shown in Eq.~(\ref{eq.eta_rev}).
Meanwhile, we derived the irreversible efficiency in Eq.~(\ref{eq.main_eta}) as the sum of the reversible efficiency and the correction representing the effect of dissipation in a unified manner.
This suggests that our result serves as a natural extension of the efficiency of the reversible heat engines to that of irreversible heat engines based on the linear response theory.

\begin{acknowledgements}
This work was supported by JSPS KAKENHI Grant Number 19K03651.
\end{acknowledgements}


\begin{references}
\bibitem{note2} This is a Stirling cycle without a regenerator. Because of the isochoric processes, this heat engine cycle has more than two heat baths.
Thus, we can relax the equality condition in Eq.~(\ref{eq.eta_max}) as far as we consider the lowest order of efficiency in the linear response regime;
in this regime, we can choose arbitrary processes as those connecting the isothermal processes because they merely contribute to the efficiency as higher-order corrections.
\end{references}
\end{document}